\documentclass[prb,showpacs,nofootinbib,twocolumn,amsmath]{revtex4}
\usepackage{hyperref}
\usepackage{bm,amsfonts, mathtools}
\usepackage{graphicx, wasysym}
\usepackage{textcomp}
\usepackage{amssymb,xcolor,latexsym,epsfig}

\newcommand{\be}{\begin{equation}}
\newcommand{\ee}{\end{equation}}
\newcommand{\beqn}{\begin{eqnarray}}
\newcommand{\eeqn}{\end{eqnarray}}

\renewcommand{\>}{\rangle}
\newcommand{\reff}[1]{(\ref{#1})}

\def\tea{\color{teal}}

\def\nn{\nonumber}

\def\mL{\mathcal L}
\def\mD{\mathcal D}
\def\Fmn{F_{\mu \nu}}

\begin{document}
\date{\today}
\vspace*{0.5cm}

\title{Larger physical volume and  bounds  on the number of matter fields in noncompact gauge theories on a lattice}
                                                                                                                                                                                                                                                                                                                                                                                                                                                                                                                                                                                                                                                                                                                                                                                                                                                                                                                                                                                                                                                                                                    
\author{F. Palumbo}
\affiliation{
INFN, Laboratori Nazionali di Frascati, 00044 Frascati, Italy} 

\begin{abstract}
  The work was  motivated  by the numerical result that in a pure  SU(2) gauge theory the ratio $R$ of the effective non-compact and compact lattice spacing  is larger than 1 and increasing with decreasing gauge coupling, as well as  the expectation that it should further increase extending  the  parameter space. This means that  with a noncompact regularization, at given number of lattice sites and comparable scaling, one can obtain a larger physical volume, whose importance for the control of size effects has  long been known. We confirm qualitatively results and expectation by a perturbative  evaluation of the effective lattice spacing in  an expansion in the Plank constant of non-compact pure SU(2) and Abelian gauge theories, but we find in addition  that $R $ reaches the maximum value of $\sqrt 2$. Including matter fields  we find that R increases (still up to $\sqrt 2$) or decreases depending on the difference between the number of scalar and spinor degrees of freedom, and there are bounds on such a difference.
 \end{abstract}

\maketitle

%%%%%%%%%%%%%%%%%%%
\section{Introduction}
 In recent years there has been an intense activity on multiflavour gauge theories in different contexts: asymptotic safety [\onlinecite{Liti}] with large number of colours and matter fields in four space-time dimensions  and condensed matter and critical phenomena in three space-time dimensions  with Abelian  [\onlinecite{Peli1,Bona1,Smis}]  or SU(2)  [\onlinecite{,Bona2}] gauge fields, especially in connection with a possible symmetry enlargement [\onlinecite{Sach}] at a second order phase transition. 
 
Although the above themes constitute in perspective one of our motivations, the present  work started in a different context, which turned out to be related to the above, motivated by the results of a numerical simulation of pure SU(2) gauge theory on a lattice: Di Carlo and Scimia [\onlinecite{DiCa}] found that, in the parameter space accessible to them, the  effective lattice spacing  in the non-compact regularization [\onlinecite{Palu1}] is  up  to 20 per cent larger than in Wilson's one [\onlinecite{Wils1}]. Moreover  they advanced the idea that the ratio of such spacings, called $R$, should further increase by extending the parameter  space. Now the need of a large physical volume was recognized long ago [\onlinecite{Sinc}], and for recent references see [\onlinecite{Part}].  But a  larger physical volume   at fixed number of sites and comparable scaling can be obtained with a larger effective spacing [\onlinecite{Scim}]. We will not tackle this issue in the present work but the computations presented are a necessary preliminary step.

So we wanted to see up to which point the DiCarlo Scimia  [\onlinecite{DiCa}] expectation would be confirmed and we  came to the conclusion, by a perturbative expansion to first order in the Plank constant, {\it  that R does indeed continue to grow but only up to $\sqrt{2}$, which puts a bound on the parameter space}.

In the course of such an investigation it has been natural to wonder how our results  would  change by inclusion of matter fields,  which is our additional motivation related to the subjects  quoted at the beginning, and we found  by the same  perturbative expansion that $R$ increases  (still up to $\sqrt 2$) or decreases depending on the difference between the number of scalar and spinor degrees of freedom, and there are bounds on such a difference.
Because we restrict ourselves to Abelian and SU(2) gauge fields, we cannot compare our results with the Veneziano limit [\onlinecite{Vene}] and then with asymptotic safety [\onlinecite{Liti}], in which both numbers of colours and matter fields are let to grow, but we hope to come back to this question in a future work. 

Let us  define what we mean by non-compact gauge theories on a lattice. In general such a name is given to regularizations  in which the kinetic part of the Lagrangian is written in terms of the (non-compact) gauge fields while the coupling with matter fields is written in terms of the unitary Wilson variables. Hoping that by our premise we can avoid any confusion, we call instead such a formulation (which does not exist for non-Abelian theories) half compact, to distinguish  it from the totally non-compact one that we are going to use, in which  the entire  Lagrangian is written in terms of non-compact fields. Such a regularization [\onlinecite{Palu1}]  was  constructed with the idea of being  closer to the continuum  since the gauge fields  live in the algebra of the gauge group (at variance with Wilson's one  [\onlinecite{Wils1}] in which they are elements of the group). Exact gauge invariance  at finite lattice spacing is enforced by means of  auxiliary fields which decouple in the continuum limit. In the SU(2) case  an evidence for confinement  was found [\onlinecite{Palu2}] comparable to that of Wilson (at the time in which the comparison was made), while  perturbative studies of the scaling properties  showed that it has the same fixed point as  Wilson's one [\onlinecite{Becc1,Bora}]. The fixed point is the same but the renormalization trajectory is different as also showed by the numerical simulation [\onlinecite{DiCa}].

As already said at present we cannot compare with the Veneziano limit [\onlinecite{Vene}] on the number  of colours and matter fields, because our results are restricted to $N_c=2$. In general  in non-compact gauge theories there are  coloured auxiliary fields which have the same coupling as the physical ones, and one neutral  auxiliary field which has a different  coupling [\onlinecite{Palu1,Palu3}]. The case of  SU(2) is simpler because there are no coloured auxiliary fields  [\onlinecite{Becc}] and the coupling of the neutral field has been shown to be a function of the gauge coupling [\onlinecite{Becc1,Bora}]. We presume that this will remain true also for $N_c>2$, but their relationship, which should  involve the number of  matter fields, is not presently known. Apart from this, extension of our results to $N_c>2$ should not present any difficulty and we plan to do it in a future work.

Our results are  also valid for  Abelian theories, which while in 4 dimensions after several studies [\onlinecite{Azco1}] have been shown to be trivial, at least coupled to scalars or to fermions with 4  flavours in the chiral limit [\onlinecite{Kogu1}], in three dimensions are truly interacting theories.

We  work in  Euclidean space, setting $c=1$. 

The work is organized in the following way. In Section II we describe for the convenience of reader and to fix the
notation the known formulation of the non-compact regularization of SU(2) and Abelian gauge theories in Cartesian and
polar form. In Section III we derive the  expression of the effective lattice spacing in four dimensions  to first
order in the Plank constant,  compare our perturbative values   with the numerical results [\onlinecite{DiCa}] and
derive the bounds mentioned  above on the number of matter fields. 
In Section IV we briefly discuss the change to three dimensions and in Section V we present our conclusions.

\section{Non-compact  gauge theories on a lattice}

In this Section we report the formulation of  the non compact regularization of the SU(2) gauge theory on a lattice.  Subsection A is devoted to its formulation in terms of Cartesian fields: This is close to the continuum and shows the role of the auxiliary field in ensuring exact gauge invariance at finite lattice spacing. In  Subsection B the interaction with matter fields is presented and in Subsection C  the formulation in terms of polar fields: This one is close to the Wilson formulation and more convenient for the perturbative calculations of the work. The case of Abelian fields can be simply derived from the non-Abelian one, and its explicit treatment can be found in Ref.[\onlinecite{Babu}].

\subsection{Cartesian fields}
The basic ingredient of the non compact regularization of SU(2) gauge theories is the quaternion
\be
D_\mu (x) =V_\mu (x)+ i\, A_{\mu a} (x)  T_a 
\label{D}
\ee
where $V_\mu $ is a real  auxiliary field, $A_{\mu a} (x) $ are the gauge fields  and the $T_a = \sigma_a/2$ are the generators of the gauge group, $\sigma_a$ 
being the Pauli matrices. We assume a convention of summation  over repeated  indices.  

In terms of this quaternion one can define the quantities
\be
F_{\mu \nu} (x) = \frac1i \left[D_\mu (x)\,D_{\nu}(x + \mu) - D_\nu (x)\,D_{\mu} (x + \nu)\right]
\ee
 
\begin{align}
\label{eq:carYM}
\mL_{\rm YM} (x) &= {\tea \frac{\beta}8} \sum_{\mu \nu} \mbox{Tr}  [\Fmn^\dag (x) \Fmn (x)] \\
                              &=  {\tea \frac{\beta}4}\, \sum_{\mu \nu}  \mbox{Tr} \Big[D^\dag_\mu (x + \nu) D^\dag_\nu (x) D_\nu (x) D_\mu (x + \nu) \nn \\
                              &\qquad  - \,D^\dag_\nu (x + \mu) D^\dag_\mu (x) D_\nu (x) D_\mu (x + \nu)\Big] \nn 
\end{align}
where $ \mbox{Tr} $ is the trace over colours.

Under the gauge  transformations
\be
g(x) = \exp (-i\,\theta_a(x) T_a) 
\label{g}
\ee
$ D_\mu (x) $ and   $F_{\mu \nu}$  transform $\it homogeneously$
\begin{align}
D'_\mu (x) &= g(x) D_\mu (x) g^\dag (x + \mu), 
\nn \\
\Fmn' (x) &= g (x)\,\Fmn (x)\,g^\dag (x + \mu + \nu) 
\end{align}
while  $\mL_{\rm YM} (x)$ is invariant. 
For small $\theta_a(x)  $
\begin{align}
A'_{\mu a} &= A_{\mu a}
- \frac12 \epsilon_{abc}A_{\mu b} \Big[ \theta_c(x+ \mu) - \theta_c \Big]
\nonumber\\
& + V_{\mu}  \Big[ \theta_a (x+ \mu) - \theta_a \Big]
\nn \\
V'_{\mu} &= V_{\mu} - \frac{1}{4}  A_{\mu a} \Big[  \theta_a(x+ \mu)- \theta_a  \Big] \,. \label{homogeneous}
\end{align}
What has been presented so far is a matrix model, caracterized by two related features: the fields of the quaternion, unlike physical gauge fields  on a latice, transform homogeneously  and instead of  discrete spacetime there appears only a four dimensional grid of points: there is no spacetime scale and then no derivative. 

One can   make contact with  physical gauge theories  in two steps. Firstly one can introduce  a bare lattice spacing "a" by means of the gauge-invariant potential
\be
\mathcal{L}_c = \frac12 \gamma^2\sum_\mu  \Big[ D_\mu^\dag D_\mu - \frac1{a^2} \Big]^2  
\ee
and define the Lagrangian 
\be
\mathcal{L} = \mathcal{L}_{YM} +  \mathcal{L}_c 
\,.
\ee
Yet there is no space-time scale, but  the minimum of the potential  does not occur  at $V_\mu=0$, but at 
\be
\bar{V}_\mu = \pm \frac1a \,.
\ee
In non perturbative calculations one has to account for both minima, what can be naturally realized in the polar representtion [\onlinecite{Becc1}]. In a perturbative calculation, however, one can expand about one of the minima (jumps between them are energetically suppressed)  introducing the fluctuations of the auxiliary field
\be  
V_\mu=\bar{V}_\mu- W_\mu  \label{fluctuation}
\ee
so that
\be
D_\mu (x) =\frac1a - W_\mu (x)+ i\, A_{\mu a} (x)  T_a \,.
\label{Deriv}
\ee
{\it This is the second step  towards a contact with physical gauge fields, as we are going to show.}
The above change of variables  does  not break gauge invariance, but it changes the homogeneous gauge fields transformations 
into  non homogeneous ones. For small $\theta_a(x)$ such transformations  can be found in Ref.[\onlinecite{Becc}] (where the group generators have a different normalization)
\begin{eqnarray}
A'_{\mu a} &=& A_{\mu a}+ \nabla_{\mu}^{+} \theta_a - \epsilon_{abc}A_{\mu b} \theta_c
\nonumber\\
& - &a  \Big( W_{\mu} \nabla_{\mu}^{+} \theta_a  + \frac12 
\epsilon_{abc}A_{\mu b} \nabla^+\theta_c \Big)
\nn \\
W'_{\mu} &= &W_{\mu} + \frac{1}{4} a A_{\mu a}\nabla_{\mu}^{+} \theta_a   \label{phys-transf}
\end{eqnarray}
where 
\be
\nabla_\mu^\pm f= \pm\,\frac1a [f(x \pm \mu ) - f(x)] 
\ee
is the ordinary forward-leftward derivative on the lattice. 
Notice that in the formal continuous limit, $a \rightarrow 0$,  $A_\mu$ transforms as the physical vector potential while  the auxiliary field   $W_\mu$ becomes invariant  and decouples as confirmed by the expression of the potential 
\begin{eqnarray}
\mathcal L_c  &=&  
\frac12 \gamma^2 \sum_\mu \Big\{ D_\mu^{\dagger} D_\mu - \frac1{a^2}\Big\}^2
\nonumber\\
&=&\frac12 \gamma^2 \sum_\mu
\Big\{ - \frac2{a^2} W_\mu + W_\mu^2 + \frac14\, A_{\mu a}A_{\mu a}\Big\}^2 \label{potential}
\end{eqnarray}
which gives to the auxiliary field $W_\mu$ the square mass $4 \gamma^2 / a^2$. 

An arbitrary gauge invariant function $ {\mathcal P} $ of the gauge fields can be added for particular purposes, but only when such a function is a polynomial is the theory strictly local. In conclusion, the total gauge fields Lagrangian is
\be
{\mathcal L}_{\rm g} = {\mathcal L}_{\rm YM} + {\mathcal L}_{\rm c} +  {\mathcal P} \,.  \label{poly}
\ee
The couplings $ \beta$ and $ \gamma$ are not independent from each other. Firstly stability of the classical action requires [\onlinecite{Becc1}] 
\be
\gamma^2 > \frac32 \beta = \frac6{g^2}  \label{cond1}
\ee
where $g$ is the gauge coupling constant.
Secondly,  perturbation theory to one loop  gives [\onlinecite{Becc1}]
\be
\gamma = \gamma_1\,\beta + \gamma_2  \,\,\, \label{cond2}
\ee
where $ \gamma_1, \gamma_2$ are independent of  $\beta$,  $\gamma_1$ is arbitrary and $\gamma_2$ can be, but it has not been evaluated. This result has been confirmed by a calculation in the Hamiltonian formalism [\onlinecite{Bora}].

 \subsection{Matter fields Lagrangians}
Let us now write down the interactions with matter fields. For a field $\phi$ transforming as
\be
\label{eq:phitra}
\phi' (x) = g (x)\,\phi (x)
\ee
one can define right/left covariant derivatives 
\begin{align}
(\mD^+_{\mu} \phi) (x) &= D_\mu (x)\,\phi(x + \mu) - \frac1a\,\phi(x) \nn\\
(\mD^-_{\mu} \phi) (x)  &=  \frac1a\,\phi(x) - D_{\mu}^\dagger (x - \mu)\,\phi(x - \mu) 
\end{align}
and the symmetric covariant derivative 
\be
(\mD_\mu \phi) (x) = \frac12\,[D_\mu (x)\,\phi(x + \mu) - D_\mu^\dagger(x - \mu) \phi(x - \mu)].
\ee
They all transform according to Eq.\eqref{eq:phitra}. 
The Lagrangian of a scalar field $\phi$ with such a transformation is
\be
{\mathcal L}_\phi = \frac12\,\sum_\mu | {\mathcal D}_\mu \phi |^2 
+ \frac12\,m_\phi^2  | \phi |^2 + {\mathcal V}(| \phi |)   \label{scalar}
\ee
where  ${\mathcal V} (|{\phi} |)$ is a  potential. We remind our convention of summation on repeated indices and we use the notation 
\be
| \phi |^2 =  \big (\phi^f \big)^* \phi^f \,.
\ee

The Lagrangian of a spinor field $\psi$ transforming according to Eq.\eqref{eq:phitra}  is
\be
{\mathcal L}_\psi = \frac{i}2\,\sum_\mu \bar{\psi} \gamma_\mu {\mathcal D}_\mu \psi + m_\psi \bar{\psi} \psi .  \label{spinor}
\ee
Such a  Lagrangian is usually called naive because it is plagued by the so called doubling problem, which  will be ignored  in the present work.

\subsection{Polar fields}
 We still denote by  $D_\mu$ the covariant derivative in terms of polar fields
 \be
D_\mu (\rho_\mu, \alpha_\mu)  = \frac1a \rho_{\mu} U_{\mu}        \label{dpolar}
 \ee
where the polar radius $\rho_\mu$ is  the  auxiliary field  now gauge invariant and 
 \be
 U_\mu = \exp( i a \,g \, \alpha_{\mu a} T_a )  \label{Wilson}
 \ee
 are the Wilson variables parametrized in terms of the angular fields $\alpha_\mu (x)$. The gauge coupling constant has been inserted in the definition of Wilson's variables in order to get rid of it in the  perturbative kinetic energy of the $\alpha$-fields in Eq.\reff{L-quadratic}.
The polar fields  are related to the Cartesian ones according to 
\begin{eqnarray}
\rho_\mu  = &&a \Big[ \Big(W_\mu - \frac1a \Big)^2+ A_\mu^2 \Big]^{ \frac12} 
\nonumber\\
U_\mu =&&  \frac a{\rho_\mu}\,\left(\frac1a - W_\mu + i\,A_\mu \right)\,.
\end{eqnarray}
Notice that $\rho_\mu  $ is not a quaternion but it is a real  function of the lattice sites.
The Yang-Mills Lagrangian density in polar form is 
\begin{align}
 {\mathcal L}_{\rm YM} (x)&= \frac{\beta}8 \sum_{\mu,\nu}  \left\{ \rho^2_{\nu}(x+\mu) \rho^2_{\mu}(x)  \right. \nn \\
                                         &\qquad\qquad \left. - \rho_{P\mu \nu} (x) \mbox{Tr} \left( U_{P\mu,\nu} (x) -a^{-4 } \right)\right\} 
\end{align}
where
\begin{align}
\rho_{P\mu \nu}(x) &= \rho_{\mu}(x) \rho_{\nu}(x+\mu) \rho_{\nu}(x)  \rho_{\mu}(x+ \nu)
\nn\\
U_{P\mu \nu} (x) &= \frac1{a^4}\,\left[1 - U_\mu^\dag (x + \nu) U_\nu^\dag (x) U_\mu (x) U_\nu (x + \mu)\right]
\end{align}
while 
\be
\mathcal L_c = \frac12 \gamma^2 \frac1{a^4}\sum_\mu \Big(\rho_\mu^2  - 1\Big)^2 \,.
\ee
 Notice   that in the limit $\gamma \rightarrow \infty$ the compact regularization is recovered.
The gauge fields partition function  in polar coordinates is
\be
T_{\rm g} = \int_0^\infty  d \rho_\mu \rho_\mu^3\int dU_\mu \exp \Big[- a^4 \sum_x  ({\mathcal L}_{\rm YM} +{\mathcal L}_c +{\mathcal P}) \Big] \label{T(rho)SU(2)}
\ee
where $dU_\mu$ is the Haar measure. The measure  $\rho_\mu^3$ can be eliminated by assuming the arbitrary function
\be
\mathcal P = \frac{\hbar}{a^4} \sum_\mu \ln \rho_\mu^2.  \label{P4} \,.
\ee
 We note that since it is not a polynomial when expressed in terms of the Cartesian fields, strict locality is lost. The partition function can now be rewritten in the form
\be
T_{\rm g} =  \frac12 \int_0^\infty  d\rho_\mu^2  \int dU_\mu \exp \left(- a^4 \sum_x {\mathcal L}_{\rm YM} +  {\mathcal L}_{\rm c} \right) \,.
\ee

%%%%%

\section{Effective spacing for SU(2) gauge theories}
In this Section we work in 4 space-time dimensions and therefore we do not discuss Abelian theories that in such a case are trivial.

As we are  going to show  the ratio $R$ between lattice  spacing in the non-compact and compact regularization   is determined by the  value of the polar field $\rho_\mu $  at the minimum of the effective potential that we will  evaluate to first order in the Plank constant $\hbar$.

For perturbative calculations the Yang-Mills Lagrangian density  can be rewritten, ignoring irrelevant terms [\onlinecite{Becc1}], as
\begin{eqnarray}
{\mathcal L}_{\rm YM} (x) &&=  \frac12 \beta a^4 \,\sum_{\mu,\nu}\rho_\nu (x) \rho_\mu (x + \nu) \rho_\mu (x) \rho_\nu (x + \mu)  \nn \\
                                        &&\times \,{\rm Tr}\,U_{P \mu\nu} (x)\,.  \label{Y-M}
\end{eqnarray}
 It is convenient to perform the change of variables
\be
\rho_\mu^2 =1 + \frac{a^2}{\gamma} t_\mu
\ee
which makes the potential $ \mathcal L_c $ quadratic
\be
\mathcal L_c = \frac12 \sum_\mu t_\mu^2   \,.
\ee
{\it Notice that $\gamma$ appears now as an inverse coupling constant in the Yang-Mills Lagrangian of Eq.\reff{Y-M}}.

The total gauge fields  Lagrangian density in the quadratic approximation  is 
\begin{eqnarray}
{\mathcal L}_g &&= \frac12  \sum_\mu t_\mu^2 +\frac12 \sum_{\mu \nu} \rho_\mu^2 \rho^2_\nu (\nabla^+_\mu \alpha_\nu - \nabla^+_\nu \alpha_\mu )^2
\nonumber\\
&& +\frac1{\xi}  \left(  \sum_\mu \rho_\mu \nabla^+_\mu \alpha_\mu  \right)^2 
+  \bar \eta \nabla_\mu^-\rho_\mu \nabla_\mu^+\eta
 \,\,\, \label{L-quadratic}
\end{eqnarray} 
where the last terms are  gauge fixing and ghost contributions [\onlinecite{Becc1}]. The gauge coupling constant disappears at this order because of the definition \reff{Wilson} of Wilson's variables.

We evaluate the effective potential following the functional method of Jackiw[\onlinecite{Jack}]. We firstly perform the shift, equal for simplicity in all directions   
\be
t_\mu \rightarrow \hat {t}+   t_\mu  
\ee
and accordingly we  rewrite $ \rho_\mu$
\be
\rho_\mu^2 = \hat{\rho}^2 \Big (1+ \frac1{\gamma}\hat{a}^2 t_\mu  \Big) 
\ee
where
\be
\hat{\rho}^2 = 1+ \frac1{\gamma} a^2  \hat t \,, \,\,\, \hat{a} = \frac{a}{ \hat{\rho}}\,.
\ee
Now the covariant derivative reads
\be
D_\mu= \frac1{\hat a} \sqrt{ 1+ \frac1{\gamma} \hat{a}^2 t_\mu }  \, \, U_\mu
\ee
showing that ${\hat a}  $ is the effective lattice spacing, namely the physical spacing, and the ratio of the spacings is 
\be
R=\frac{1}{ \hat {\rho}}\,.
\ee

\subsection{Evaluation of the effective potential} 
For the  evaluation of the effective potential it is convenient to perform the further change of variables
\be
t_\mu = \frac1{a^2} s_\mu \,, \,\,\,\,\, s_\mu \rightarrow \hat{s}+ s_\mu\,.
\ee
Then following Jackiw [\onlinecite{Jack}] we define the shifted  Lagrangian
 \begin{eqnarray}
 \hat{{\mathcal L}}_g &&= \frac12  \sum_\mu ( s_\mu^2 + \hat{s}_\mu^2  )+\frac1{2} \hat{\rho}^4\sum_{\mu \nu}  (\nabla^+_\mu \alpha_\nu - \nabla^+_\nu \alpha_\mu )^2
 \nonumber\\
 &&+\frac1{\xi}  \hat{\rho}^2 \Big(\sum_\mu  \nabla^+_\mu \alpha_\mu \Big)^2
 +\hat{\rho} \, \bar \eta \sum_\mu\nabla_\mu^- \nabla_\mu^+\eta
\end{eqnarray} 
obtained by retaining only quadratic terms  in all the fields fluctuations and neglecting the linear ones. The partition function can now be written
\be
T_{\rm g} =  \frac12 \int_{-\gamma}^\infty  ds_\mu  \int dU_\mu 
\exp \left(- a^4 \sum_x \rm  \hat{{\mathcal L}}_g  \right) 
\ee
where $dU_\mu $ is the Haar measure.
It follows that the classical gauge fields effective potential is
\be
\mathcal{V}_g^{(0)} = 2 \, \hat s^2\,.
\ee
 It is what remains of the potential \reff{potential} after all our manipulations. Its minimum occurs at $\hat s=0$, namely at the bare lattice spacing.
We then  evaluate the partition function to order  $ \hbar $. 
Integration over $s_\mu$ can be ignored because it does not change the dependence on $ \hat{s} $ and then on $ \hat{\rho} $ of the effective potential.

After  Fourier transformation we get the action 
\be
a^4 \sum_x  \hat{{\mathcal L}}_g = \frac1{N^4} \sum_p \tilde {\alpha}^*_{\mu c}(p) M_{\mu \nu}(p) \, \alpha_{\nu c}(p)
\ee
where $N$ is the number of lattice spacings in one direction and the matrix 
\be
M_{\mu \nu}(p) = \delta_{\mu \nu} p^2 \hat{\rho}^4 - \Big( \hat{\rho}^4 - \frac1{\xi}  \hat{\rho}^2\Big) p_\mu p_\nu 
\ee
has  determinant 
\be
\det M(p)= \frac1{\xi} (p^2 )^4  ( \hat{\rho}^2)^7 \,.   \label{det}
\ee
In the same way one obtains the ghost contribution for each momentum and colour
\be
(\hat \rho^2) ^{\frac12 } \,.
\ee
Therefore ignoring  irrelevant factors
\be
T_g= ( \hat \rho^2)^{-9 N^4} \,.
\ee
Following Jackiw [\onlinecite{Jack}], the  contribution of order $\hbar$ to the effective potential  can be obtained by dividing the effective action by the space-time volume 
 \be
\mathcal{V}_g^{(1)}= 9 \frac{\hbar}{a^4} \,   \ln \hat \rho^2 
\ee
so that the total effective potential  turns out to be 
\be
\mathcal{V}_g = 2 \frac{\gamma^2 }{a^4} 
\Big[ ( \hat{\rho}^2 -1)^2 + \frac12 \, \sigma_g \, \ln \hat{ \rho}^2 \Big] 
\ee
where 
\be
\sigma_g= 9 \frac{\hbar}{\gamma^2} \,.
\ee
{\it At first sight it might be surprising that the gauge coupling constant does not appear, but we should remember that at variance with standard cases, here the shifted field is the auxiliary field, whose coupling constant  is indeed $1/ \gamma$}. We should also remember  that according to Eq.\reff{cond2} $\gamma$ is a function of $\beta$.

The coupling constant $\frac1{\xi} $ of the gauge fixing disappeared, but obviously this does not mean per se' that  the result is gauge independent. There are several investigations with conclusions that physical results derived from the effective potential  are gauge independent [\onlinecite{Niel}]. The ratio of lattice spacings is not a physical quantity, but in the following subsection we will extend our analysis including the coupling with matter fields, and then we will derive bounds on their number which are physical results. Because  our procedure is not standard however,  it is not obvious to us that  the quoted results [\onlinecite{Niel}] are valid also in the present case. We leave such a question for a future investigation.

The effective potential  diverges to $ - \infty$ for  $ \hat{\rho}^2 \rightarrow 0$ but the gauge fields Lagrangian  is bounded from below [\onlinecite{Becc1}], so that such negative divergence is not a problem of the regularization but of the perturbative expansion that cannot be trusted in this region. We do not think  that it should cast doubts on the validity of the correction evaluated at finite  lattice spacing, and therefore we think that it can be ignored in the present context. All the above refers to a pure SU(2) theory, and the situation can be changed by matter fields as shown in Subsection C.

 The first derivative of the effective potential
\be
\frac{\partial \mathcal{V}_g}{\partial  \hat{\rho}^2} = 4 \frac{\gamma^2  }{a^4} \frac{1}{\hat{\rho}^2}
\Big[ ( \hat{\rho}^2 )^2 - \hat{\rho}^2 +\frac14 \sigma_g\,   \Big] 
\ee
vanishes at
\be
\hat{\rho}^2_{\pm} = \frac12 \Big[ 1\pm  \sqrt{ 1-  \sigma_g} \Big]   \label{rosina}
\ee
so that the effective potential  takes a  maximum  at $\hat{\rho}_- $, and a local minimum at $\hat{\rho}_+ $ with the following exceptions. For   $\sigma_g = 1$, $\hat{\rho}_{\pm} = 1/ \sqrt 2$ is neither a maximum nor a minimum but an inflection. For  $\sigma_g = 0$,  the Wilson limit, the minimum  occurs at  $\hat{\rho}=1$, namely $ \hat{a} = a$.

Otherwise the roots are real under the condition on $\gamma$
 \be
\frac{\gamma^2}{\hbar} > 9 \label{cond3}
\ee
to be compared with the condition \reff{cond1}. Under such a condition, since the  theory is defined at the minimum of the effective potential
\be
1 < R= \frac{\hat{a}}{a} = \frac{1}{ \hat{\rho}_+}  < \sqrt 2  \,.  \label{R}
\ee

\subsection{Comparison with the  numerical simulation of Di Carlo and Scimia}

 Di Carlo and Scimia [\onlinecite{DiCa}] made a plot of  the ratio of two hadron masses as a function of $ 1/ \gamma$ and $\beta$. The resulting surface has a valley starting at the Wilson point $1/\gamma=0$ that they regarded as a scaling window and called  the scaling valley. {\it The bottom of this valley at given $1/ \gamma$ determines this parameter  as a function of $\beta$, a function which was found consistent with the perturbative one \reff{cond2}}. This allowed them to make a plot (in their Fig.7) of the  lattice spacing $\hat{a}_{\text{num}}$, which is a function of both $\gamma$ and $\beta$, as a function of  $\gamma$ only, which is directly comparable to our result which comes out naturally as a function of $\gamma$. 
 
 Their plot is done  in the interval  $0.0 \le  1/ \gamma  \le 0.18$. We now compare some values of $R$  extracted from the plot, which we call $R_{\text{num}}$, with ours evaluated according to Eq. \reff{R} 
\begin{eqnarray}
\frac1{\gamma } &=& 0.00 \,\,\,\,\, R=1     \,\, \,\, \,\,\,\,\,\,\,\,\,\,\     R_{\text{num}}= 1.00 
\nonumber\\
\frac1{\gamma } &= &0.06\,\,\,\,\,  R =1.00       \,\,\,\,\,\,\,\,      R_{\text{num}}= 1.02 
\nonumber\\
\frac1{\gamma } &= &0.12 \,\,\,\,  R =1.02      \,\,\,\,\,\,\,\,      R_{\text{num}}= 1.07
\nonumber\\
\frac1{\gamma} &=&0.18 \,\,\, \,\,  R =1.04      \,\,\,\,\,\,\,\,     R_{\text{num}}= 1.15 
\nonumber\\
\frac1{\gamma} &=&0.24\,\,\, \,\,\,  R =1.09
\nonumber\\
\frac1{\gamma} &=&0.30 \,\, \,\,\,  R =1.18
\nonumber\\
\frac1{\gamma} &\rightarrow & \,\frac13 \,\, \,\,\, \,\,\, \,\, R \rightarrow \sqrt 2 \,.
\end{eqnarray}
We see that the agreement is only  qualitative, there being a factor four of discrepancy. However we confirm the  expectation of Di Carlo and Scimia that  $R$ should have  increased extending the range of $\gamma$, which was the original motivation of the present work, but we find the additional result that in pure SU(2) $\gamma$ is bound according to Eq.\reff{cond3}.

\subsection{Inclusion of  matter fields}
Inclusion  of matter fields is   straightforward. From Eqs.\reff{scalar},\reff{spinor},  using the definition \reff{dpolar} of covariant derivative and retaining only  quadratic fluctuations  we get the matter fields shifted Lagrangians
\begin{eqnarray}
\hat{{\mathcal L}}_\phi &&= \frac12  \hat{\rho}^2 \sum_\mu  | \nabla_\mu \phi |^2 
+ \frac12\,m_\phi^2  | \phi |^2
\nonumber\\
\hat{{\mathcal L}}_\psi &&= \frac{i}2\,\hat{\rho}\sum_\mu \bar{\psi} \gamma_\mu \nabla_\mu \psi + m_\psi \bar{\psi} \psi .     
\end{eqnarray}
 As for the pure SU(2)  theory and for the same reason, self interactions and Yukawa couplings would not appear in the effective potential, so we omit them.

Assuming for simplicity matter fields to be massless, to first order in the Plank mass they give the effective potential  the contribution
\be
\mathcal{V}_{\text{mat}} = \frac12 \frac{\hbar}{a^4} (\nu_s - \nu_f)  \ln  \hat{\rho}^2
\ee
where  $\nu_s , \nu_f $ are the numbers of scalars and fermion degrees of freedom respectively: $\nu_s$ is  the  colour times flavour dimension of the scalars representations, $\nu_f$  the colour times flavour times  Dirac matrices dimension of the fermion representation. The latter one is 4  in four  space-time dimensions, but 4 or 2 in three space-time dimensions depending on whether parity is conserved or not. 

In conclusion now the total effective potential reads
\be
\mathcal{V}= 2 \frac{\gamma^2 }{a^4} 
\Big[ ( \hat{\rho}^2 -1)^2 + \frac12 \, \sigma \, \ln \hat{ \rho}^2 \Big] 
\ee
where 
\be
\sigma= \frac{1}{ 2 } (18 + \nu_s - \nu_f) \frac{\hbar}{\gamma^2}\,.
\ee 
If $\sigma >0$ the equations derived for pure gauge theories remain valid after replacing $ \sigma_g$ by $\sigma$ 
 \be
0<  18 + \nu_s -  \nu_f  < 2 \frac{\gamma^2 }{\hbar}= \frac2{\hbar}(\gamma_1 \beta + \gamma_2)^2 \,.
\ee
The first unequality is a bound on the number of degrees of freedom of all the fields, independent of the coupling constants, while  the second one, which replaces \reff{cond3},  involves the parameter $\gamma$ and then the inverse coupling $\beta$. It can be regarded as a further bound on the number of fields at given $\gamma$, or as a condition on this parameter for given number of fields. 

A  completely different situation occurs if $ \sigma<0$
\be
 18 + \nu_s -  \nu_f <0 \,.
\ee
In such a case the  effective potential for  $\hat{ \rho} \rightarrow 0 $ diverges to $+ \infty, \,  \hat{ \rho}_{-} $ becomes imaginary and must be discarded, {\it  $ \hat{ \rho}_{+} >1$ becomes an absolute minimum so that $R<1$, namely the effective lattice spacing is smaller than the bare one.}

\section{Three dimensions}
 All the above concerns the theory in four space-time dimensions. 
In order to evaluate the effective potential in three dimensions it is  necessary to change the function $\mathcal {P}$ \reff{P4}, chosen to get rid of the measure of integration on the radial field in 4 dimensions  which now must be 
 \be
 \mathcal{P}= \frac1{a^3} \hbar \sum_\mu \ln \rho_\mu \,.
 \ee
 Then for SU(2) gauge theories  we look at the expression of the determinant  of the gauge fields matrix \reff{det} for each momentum and colour: reducing the space-time  dimensions by one  it becomes
 \be
 \det M(p)= \frac1{\xi}G (p^2 )^3  ( \hat{\rho}^2)^5 
\ee
so that all the results for three dimensions can be obtained by replacing $\sigma_g$ according to
\be
\sigma_g^{(3)}= 6 \frac{\hbar}{\gamma^2} \,.
\ee
Concerning Abelian gauge theories  we must remember   that while in four dimensions they have been shown to be trivial [\onlinecite{Azco1},\onlinecite{Kogu1}], in three dimensions they are  truly interacting and largely used in condensed matter physics  [\onlinecite{Peli1,Bona1,Smis}], as reminded in the  Introduction. 

 Gauge transformations as well as Lagrangian density can be  derived immediately and can be found  in Ref.[\onlinecite{Babu}]. Then in the evaluation of the shifted Lagrangian one has to omit  the ghost contribution and in the evaluation of the effective potential one must omit the factor due to colours, with the result
 \be
 ( \sigma_g^{(3)})_{Abelian}= \frac52 \frac{\hbar}{\gamma^2}\,.
 \ee

%%%%%

\section{Conclusions}

We  confirmed qualitatively to order $\hbar$  of perturbation theory the numerical results [\onlinecite{DiCa}]  that $R$, the ratio between  lattice spacings in the  non-compact and compact regularizations of the pure SU(2) gauge theory is larger than 1.  We confirmed also the expectation that $R$ should further increase with decreasing  $\gamma$, but we found that $\gamma$ has a lower bound, and for this value R reaches its maximum $R = \sqrt 2$. 

Of  course quantitatively we should trust the numerical simulations more than the perturbative results. Therefore, if the factor 4 between $R_{num}$ and {R} persists up to the minimum value of $\gamma$, we might expect the maximum of $R_ {num}$ to be of the order of 5. 

We extended the investigation including  massless matter fields and showed  that  the ratio $R$  increases (still up to $\sqrt 2$) or decreases depending on the difference between the number of scalar and spinor degrees of freedom, and there are bounds on such a difference. 

These bounds might be relevant to  the other theme which occurred as a second motivation of the work, namely ultraviolet fixed points and the related Veneziano limit [\onlinecite{Vene}], but in order to proceed in this direction two further steps are needed. The first one is the extension to large number of colours. In $ SU ( N_c  \> 2 ) $  in addition to the neutral  auxiliary field there are coloured auxiliary fields with the same coupling constant as the physical ones [\onlinecite{Palu1,Palu3}] which will contribute to the effective potential. 
The second step is the extension of the relationship \reff{cond2} to  large number of colours and  matter fields. We hope to come back to these issues in a future work.

The above results are also valid in three dimensions where they are relevant to condensed matter physics.

 At last a  comment about Abelian gauge theories. One might wonder  whether in four dimensions, due to the absence of  the relationship \reff{cond2} between $\beta$ and $\gamma$ which remain two independent couplings the noncompact regularization might define a theory different from the standard one, with a continuum nontrivial limit. This seems however very unlikely. Indeed the non-compact theory is renormalizable by power counting,  and  integration on the auxiliary field [\onlinecite{Babu}], although in a perturbative regime,  yields only irrelevant contributions.

%++++++++++++++++++++++

%%%%%%%%%%%%%%%%%%%

\section*{Acknowledgements}

\appendix
I am grateful to V. Azcoiti and G. Di Carlo for a most helpful correspondence.

%%%%%%%%%%%%%%%%%%%


\newpage
 
\begin{thebibliography}{99}

 \bibitem{Liti}
  K.G. Wilson, Renormalization Group and Critical Phenomena. I. Renormalization Group and the Kadanoff Scaling Picture, Phys.Rev. B4 (1971) 3174;
 K.G. Wilson,  Renormalization Group and Critical Phenomena. II. Phase Space Cell Analysis of critical Behaviour, Phys.Rev. B4 (1971) 3184;
 S. Weinberg, in General relativity: An Einstein centenary survey, ed. S.W. Hawking and W. Israel, 790 (1979); a short review on the subject can be found in
D.F. Litim and F. Sannino, Asymptotic safety guaranteed J.High Energy Phys. 12 (2014) 178;
A.D. Bond and D.F. Litim, More asymptotic safety guaranteed; Phys. Rev. D97 (2018) 085008; A.V. Bednyakov and A.I. Mukhaeva, Asymptotic safety in the Litim-Sannino model at four loops, Phys. Rev. 109 (2024) 065030; T. Steudtner, Effective potential  and vacuum stability in the Litim-Sannino model, J. High Energy Phys. 05 (2024)149

\bibitem{Peli1} 
 A. Pelissetto and E. Vicari, Three-dimensional ferromagnetic CP$^{(N-1)}$ models, 
 Phys. Rev. E \textbf{100}, 22122 (2019);
  Multicomponent compact Abelian-Higgs lattice models, 
 Phys. Rev. E \textbf{100}, 042134 (2019);
 Large-$N$ behaviour of three-dimensional lattice CP$^{N-1}$ models, 
 J. Stat. Mech. \textbf{2003}, 033209 (2020).
 
 \bibitem{Bona1} 
 O.I. Motrunich and A. Vishwanath, 
 Comparative study of Higgs transition in one-component and two-components lattice superconductor models, 
 arXiv:0805.1494; 
  C. Bonati, A. Pelissetto, and E. Vicari, 
 Lattice Abelian-Higgs model with noncompact gauge fields, 
 Phys. Rev. B \textbf{103}, 085104 (2021); ibid.,
 Noncompact lattice Higgs model with Abelian discrete gauge groups: Phase diagram and gauge symmetry enlargement, 
 Phys. Rev. B \textbf{107}, 035106 (2023). 

\bibitem{Smis} 
 J. Smiseth, E. Smorgav, F.S. Nogueira, J. Hove, and A. Sudbo, 
 Phase structure of (2+1)-dimensional compact lattice gauge theories and the transition from Mott insulator to fractionalized insulators,  Phys. Rev. B \textbf{67}, 205104 (2003);
 C. Bonati, A. Pelissetto, and E. Vicari, 
 Higher-charge three-dimensional compact  lattice Abelian-Higgs models, 
 Phys. Rev. E \textbf{102}, 062151 (2020).

 \bibitem{Bona2} 
 A.B. Kuklov, M. Matsumoto, N.V. Prokof'ev, B.V. Svistunov, and M. Troyer, 
 Deconfined criticality: generic first-order transition in the SU(2) symmetry case, 
 Phys. Rev. Lett. \textbf{101}, 050405 (2008);
 S. Sachdev, H.D. Scammell, M.S. Scheurer and G. Tarnopolsky, 
 Gauge theories for cuprates near optimal doping, 
 Phys. Rev. B \textbf{99}, 054516 (2019); 
 H.D. Scammell, K. Patekar, M.S. Scheurer, and S. Sachdev, 
 Phases of SU(2) gauge theory with multiple adjoint Higgs fields in 2+1 dimensions, 
 Phys. Rev. B \textbf{101}, 205124 (2020);
 C. Bonati, A. Franchi, A. Pelissetto, and E. Vicari, 
 Three-dimensional lattice SU($N_c$) gauge theories with multiflavor scalar fields in the adjoint representation, 
 Phys. Rev. B \textbf{104}, 115166 (2021).

\bibitem{Sach} 
 S. Sachdev, 
 Topological order, emergent gauge fields and Fermi surface reconstruction, 
 Rep. Prog. Phys. \textbf{82}, 014001 (2019).

 \bibitem{DiCa} 
 G. Di Carlo and R. Scimia, 
 Numerical study of the scaling properties of SU(2) lattice gauge theory in Palumbo non compact regularization, 
 Phys. Rev. D \textbf{63}, 094501 (2001).
 
 \bibitem{Palu1} 
 F. Palumbo, 
 Noncompact gauge fields on a lattice, 
 Phys. Lett. B \textbf{244}, 55 (1990).
 
\bibitem{Wils1} 
 K.G. Wilson, 
 Confinement of quarks, 
 Phys. Rev. D \textbf{10}, 2445 (1974).
 
 \bibitem{Sinc}
 D. K. Sinclair, Nucl. Phys. B (Proc. Suppl.) 94 (1996) 112; R. Petronzio, Nucl. Phys. B (Proc. Suppl.) 83-84 (2000) 136; L. Lellouch, Nucl. Phys. B (Proc. Suppl.) 94 (2001)142 
 
 \bibitem{Part}
 Particle Data Group, Lattice Quantum Chromodynamics (2024)

 \bibitem{Scim}
 G. Di Carlo, F. Palumbo and R. Scimia, Larger physical volume with a noncompact lattice regularization of SU(N) theories, Nucl. Phys. B (Proc. Suppl.) 106 (2002) 823
 
 \bibitem{Vene}
 G. Veneziano,Nucl. Phys. B159 (1979)213

\bibitem{Palu2}
 F. Palumbo, M.I. Polikarpov, and A.I. Veselov, 
 Confinement in non compact non Abelian gauge theories on a lattice, 
 Phys. Lett. B \textbf{297}, 171 (2002)
 
\bibitem{Becc1}  
 C.M. Becchi and F. Palumbo, 
 Noncompact gauge theories on a lattice: perturbative study of the scaling properties, 
 Nucl. Phys. B \textbf{388}, 595 (1992). 

 \bibitem{Bora}
 B. Borasoy, W. Kramer, and D. Schutte, 
 Application of the Hamiltonian formulation of Palumbo's new lattice Yang-Mills theory, 
 Phys. Rev. D \textbf{53}, 2599 (1996);
 B. Dickmann, D. Schutte, and H. Kroger, 
 Hamiltonian formulation of Palumbo's non compact lattice gauge theory, 
 Phys. Rev. D \textbf{49}, 3589 (1994). 
  
 \bibitem{Palu3} 
 F. Palumbo and R. Scimia, 
 Non compact gauge fields on a lattice: SU(N) theories, 
 Phys. Rev. D \textbf{65}, 074509 (2002).

 \bibitem{Becc}
 C.M. Becchi and F. Palumbo, 
 Compact and non compact gauge theories on a lattice, Phys. Rev. D \textbf{44}, R946 (1991).

\bibitem{Azco1} 
 V. Azcoiti, G. Di Carlo, and A.F. Grillo, 
 New proposal for including dynamical fermions in lattice gauge theories: The compact-QED case, 
 Phys. Rev. Lett. \textbf{65}, 2239 (1990); 
  V. Azcoiti, A. Cruz, G. Di Carlo, A.F. Grillo, and A. Vladikas, 
 Simulating lattice fermions by microcanonically averaging out the nonlocal dependence of the fermionic action,
 Phys. Rev. D \textbf{43}, 3487 (1991);
  V. Azcoiti, G. Di Carlo, and A.F. Grillo, 
 A new approach to non-compact lattice QED with light fermions,
 Int. J. Mod. Phys. A \textbf{8}, 4235 (1993);
 M. Baig, H. Fort, S. Kim, J.B. Kogut, and D.K. Sinclair, 
 Phys. Rev. D \textbf{48}, R2385 (1993);
   M. Baig, H. Fort, J.B. Kogut, and S. Kim, 
 Phys. Rev. D \textbf{51}, 5216 (1995).

\bibitem{Kogu1} 
 J.B. Kogut and E. Dagotto, 
  A supercomputer study of strongly coupled QED, 
  Phys. Rev. Lett. \textbf{59}, 617 (1987);
   E. Dagotto and J.B. Kogut,
 Study of compact QED with light fermions,
 Nucl. Phys. B \textbf{295}, 123 (1988);
 S. Kim, J. B. Kogut, and M.-P. Lombardo, 
 On the triviality of textbook quantum electrodynamics, 
 Phys. Lett. B \textbf{502}, 345 (2001);
 Gauged Nambu--Jona-Lasinio studies of the triviality of quantum electrodynamics,
 Phys. Rev. D \textbf{65}, 054015 (2002).

 \bibitem{Babu}
 D. Babusci and F. Palumbo, Modified Abelian and SU(2) Wilson theories on a lattice from a noncompact regularization,
 Phys. Rev. D 110 (2024) 094511

 \bibitem{Jack}
R. Jackiw, Functional evaluation of the effective potential, Phys. Rev. D 9 (1974) 1686;
L. Dolan and R. Jackiw, Gauge-invariant signal for gauge-symmetry breaking,  Phys. Rev. D 9 (1974) 2904

\bibitem{Niel}
N.K.Nielsen,On the gauge dependence of spontaneous symmetry breaking in gauge theories, Nucl. Phys. B 101 (1975) 173; R. Fukuda and T. Kugo, Gauge invariance in the effective action and potential, Phys. Rev. D13 (1976) 3469; L.J.R. Aitchison and C. M. Fraser, Gauge invariance and the effective potential, Annals Phys. 156 (1984) 1 
 
 \end{thebibliography}
\end{document}